\documentclass{article}
\usepackage{amsmath, amssymb, amsfonts}
\title{Rindler-type geometry inside a black hole} 
\author{Hristu Culetu, \\Ovidius University, Dept.of Physics, \\B-dul Mamaia 124, 900527 Constanta, Romania, \\e-mail : hculetu@yahoo.com}

\begin{document}
\numberwithin{equation}{section}
\pagenumbering{arabic}
\maketitle
\newcommand{\fv}{\boldsymbol{f}}
\newcommand{\tv}{\boldsymbol{t}}
\newcommand{\gv}{\boldsymbol{g}}
\newcommand{\OV}{\boldsymbol{O}}
\newcommand{\wv}{\boldsymbol{w}}
\newcommand{\WV}{\boldsymbol{W}}
\newcommand{\NV}{\boldsymbol{N}}
\newcommand{\hv}{\boldsymbol{h}}
\newcommand{\yv}{\boldsymbol{y}}
\newcommand{\RE}{\textrm{Re}}
\newcommand{\IM}{\textrm{Im}}
\newcommand{\rot}{\textrm{rot}}
\newcommand{\dv}{\boldsymbol{d}}
\newcommand{\grad}{\textrm{grad}}
\newcommand{\Tr}{\textrm{Tr}}
\newcommand{\ua}{\uparrow}
\newcommand{\da}{\downarrow}
\newcommand{\ct}{\textrm{const}}
\newcommand{\xv}{\boldsymbol{x}}
\newcommand{\mv}{\boldsymbol{m}}
\newcommand{\rv}{\boldsymbol{r}}
\newcommand{\kv}{\boldsymbol{k}}
\newcommand{\VE}{\boldsymbol{V}}
\newcommand{\sv}{\boldsymbol{s}}
\newcommand{\RV}{\boldsymbol{R}}
\newcommand{\pv}{\boldsymbol{p}}
\newcommand{\PV}{\boldsymbol{P}}
\newcommand{\EV}{\boldsymbol{E}}
\newcommand{\DV}{\boldsymbol{D}}
\newcommand{\BV}{\boldsymbol{B}}
\newcommand{\HV}{\boldsymbol{H}}
\newcommand{\MV}{\boldsymbol{M}}
\newcommand{\be}{\begin{equation}}
\newcommand{\ee}{\end{equation}}
\newcommand{\ba}{\begin{eqnarray}}
\newcommand{\ea}{\end{eqnarray}}
\newcommand{\bq}{\begin{eqnarray*}}
\newcommand{\eq}{\end{eqnarray*}}
\newcommand{\pa}{\partial}
\newcommand{\f}{\frac}
\newcommand{\FV}{\boldsymbol{F}}
\newcommand{\ve}{\boldsymbol{v}}
\newcommand{\AV}{\boldsymbol{A}}
\newcommand{\jv}{\boldsymbol{j}}
\newcommand{\LV}{\boldsymbol{L}}
\newcommand{\SV}{\boldsymbol{S}}
\newcommand{\av}{\boldsymbol{a}}
\newcommand{\qv}{\boldsymbol{q}}
\newcommand{\QV}{\boldsymbol{Q}}
\newcommand{\ev}{\boldsymbol{e}}
\newcommand{\uv}{\boldsymbol{u}}
\newcommand{\KV}{\boldsymbol{K}}
\newcommand{\ro}{\boldsymbol{\rho}}
\newcommand{\si}{\boldsymbol{\sigma}}
\newcommand{\thv}{\boldsymbol{\theta}}
\newcommand{\bv}{\boldsymbol{b}}
\newcommand{\JV}{\boldsymbol{J}}
\newcommand{\nv}{\boldsymbol{n}}
\newcommand{\lv}{\boldsymbol{l}}
\newcommand{\om}{\boldsymbol{\omega}}
\newcommand{\Om}{\boldsymbol{\Omega}}
\newcommand{\Piv}{\boldsymbol{\Pi}}
\newcommand{\UV}{\boldsymbol{U}}
\newcommand{\iv}{\boldsymbol{i}}
\newcommand{\nuv}{\boldsymbol{\nu}}
\newcommand{\muv}{\boldsymbol{\mu}}
\newcommand{\lm}{\boldsymbol{\lambda}}
\newcommand{\Lm}{\boldsymbol{\Lambda}}
\newcommand{\opsi}{\overline{\psi}}
\renewcommand{\tan}{\textrm{tg}}
\renewcommand{\cot}{\textrm{ctg}}
\renewcommand{\sinh}{\textrm{sh}}
\renewcommand{\cosh}{\textrm{ch}}
\renewcommand{\tanh}{\textrm{th}}
\renewcommand{\coth}{\textrm{cth}}

\begin{abstract}
 An anisotropic fluid with positive energy density and negative pressures is proposed in the black hole interior. The gravitational field is constant everywhere inside and is given by the horizon surface gravity. Even though the geometry is of Rindler type it is curved because of the spherical symmetry used and is singular at the origin. The Israel junction conditions are studied on the stretched horizon instead of using a matching process on the null surface $r = 2M$. The entropy of any inner sphere is mass independent, maximally packed and unaffected by the outer layers. 
 
\end{abstract}

\section{Introduction}
  A wide variety of current data supports the view that the matter content of the Universe consists of two basic components, namely dark matter (DM) and dark energy (DE) with ordinary matter playing a minor role. The nature and composition of DM and DE is not at all understood.
  
  As Mannheim \cite{PM1} has noticed, what is disturbing is the \textit{ad hoc} way in which DM is actually introduced. The DE problem is even more severe (its composition and structure is as mysterious as that of DM). The entire motivation for the existence of DM and DE is based on their validity at all distance scales of the standard Newton-Einstein gravitational theory \cite{PM1}.
         
 Recently Grumiller \cite{DG1} proposed a paradigm for gravitation at large distances. His metric generates a new Rindler acceleration term in a spherically symmetric situation, just as Mannheim stated before in his conformally invariant model of cosmology. The Rindler constant acceleration may depend on the scale of the system under consideration and becomes important at large distances from the source. 
 
  Throughout the Letter we consider geometrical units $G = c = \hbar = k_{B} = 1$.
  
  \section{Spherical Rindler-type metric} 
  We used in \cite{HC5} Grumiller's metric without the mass term \footnote{It is worth noting that the corresponding metric with the mass term, namely $g_{tt} = - g_{rr}^{-1} = -1 + 2m/r + 2ar$ has motivated Mannheim \cite{PM2} and Grumiller on their studies on the galactic rotation curves. Mannheim's constant $\gamma_{0}$ is related to the critical MOND acceleration $a_{0} = \gamma_{0} \approx 10^{-8} cm/s^{2}$. In Mannheim's view, the linearly rising potential shows that a local matter distribution can actually have a global effect at infinity and gravity theory becomes global}.
   \begin{equation}
  ds^{2} = -(1 - 2ar) dt^{2} + (1 - 2ar)^{-1} dr^{2} + r^{2} d \Omega^{2}
 \label{2.1}
 \end{equation}
as a geometry valid in the interior of a relativistic star. In (2.1), $a$ is the ''Rindler'' acceleration and $d \Omega^{2}$  stands for the metric on the unit 2-sphere. To be a solution of Einstein's equations $G_{ab} = 8\pi T_{ab}$, one shows that a stress tensor is needed on its r.h.s. , namely
\begin{equation}
 T_{~t}^{t} = - \rho = - \frac{a}{2 \pi r},~~~p_{r} = T_{~r}^{r} = - \rho,~~~T^{\theta}_{~\theta} = T^{\phi}_{~\phi} = p_{\bot} = \frac{1}{2} p_{r}
 \label{2.2}
 \end{equation}
 where $\rho$ is the energy density of the anisotropic fluid, $p_{r}$ is the radial pressure and $p_{\bot}$ are the tangential pressures. It is worth noting that the anisotropic fluid is comoving with the accelerated observer (the stress tensor is diagonal). We chose $a > 0$ in order for the metric (2.1) to possess a Rindler horizon. We ask in this paper whether the linear Rindler-type potential may be applied within a gravitational source (specifically, a black hole).
 
 Let us write down the general expression for the energy-momentum tensor for an anisotropic fluid \cite{MBC, GE}, without heat flux
 \begin{equation}
T_{~a}^{b} = (\rho + p) u_{a} u^{b} + p \delta_{~a}^{b}+ \pi_{~a}^{b}.
\label{2.3}
\end{equation}
 In (2.3), the Latin indices $a$ and $b$ span ($t, r, \theta, \phi$) coordinates, $u^{a}$ is the timelike velocity vector of the fluid, with the components $u^{a} = (1/\sqrt{1 - 2ar}, 0, 0, 0)$, $\pi_{~a}^{b}$ is the diagonal anisotropic stress tensor (associated with viscosity effects), with $\pi_{ab} = \pi_{ba},~\pi^{a}_{~a} = 0$ and $\pi_{ab}u^{a} = 0$. The anisotropy is rooted from the fact that the radial and tangential pressures are not equal. Following Mannheim et al. \cite{MBC}, we define $\pi^{\theta}_{~\theta} = \pi^{\phi}_{~\phi} = -q$ and, therefore, $\pi^{r}_{~r} = 2q$. The components of $T_{~a}^{b}$ acquire now the form \cite{MBC}
 \begin{equation}
 T_{~t}^{t} = p + 2q = T_{~r}^{r},~~~T^{\theta}_{~\theta} = T^{\phi}_{~\phi} = p - q.  
 \label{2.4}
 \end{equation}
In order Eqs. (2.2) to be in accordance with (2.4), we must have $\rho = - p_{r} = -2 p_{\bot} = - p - 2q$ and, therefore, $p = 4q$. The anisotropic stress tensor becomes now $\pi_{~a}^{b} = (0,~-a/6\pi r,~a/12\pi r,~a/12\pi r)$.

 We conjecture that the previous model may be applied inside a black hole (BH). Our choice is based on a well-known property of the Schwarzschild line element: close to the event horizon the geometry is of Rindler type. That means one is easier to fulfill the first Israel junction condition at $r = 2M$. Moreover, in our view we may not extend the exterior Schwarzschild geometry beyond the horizon because it has been initially obtained using a Newtonian approximation at $r >> 2M$. In addition, the signature flip at the event horizon is avoided. We shall see later that the above recipe leads to a model for the BH interior very similar with the Davidson-Gurwich prescription \cite{DG2}.

\section{Black hole interior geometry}
 Take now an observer located at some distance $r$ from the BH center. The gravitational field there depends only on the mass $m(r)$ up to the radius $r$, i.e. it is given by \cite{SW}
 \begin{equation}
\label{3.1}
 m(r) =  \int^{r}_{0}{4 \pi r^{2} \sqrt{-g_{tt}~ g_{rr}}~ \rho(r) dr = ar^{2}}. 
\end{equation}
 (The same result would have been obtained by using the Padmanabhan formula  \cite{TP1}). Therefore, one may remove the mass ''above'' the observer and the metric becomes ''of  Schwarzschild type''. This interpretation is in accordance with the result for $A(r)$ obtained by Mannheim \cite{PM1} in his study on anisotropic fluids in Robertson-Walker universe
  \begin{equation}
  A^{-1}(r) = 1 - \frac{2m(r)}{r},
\label{3.2}
\end{equation}
  where $A(r) = g_{rr}$. It yields in our situation $A(r) = 1/(1 - 2ar)$, while for the time-time component of the metric tensor we could make use of \cite{PM1} 
    \begin{equation}
    \frac{1}{B} \frac{dB}{dr} = \frac{2}{r} \frac{m(r) + 4\pi r^{3}p_{r}}{r - 2m(r)}
\label{3.3}
\end{equation}
with $B = - g_{tt}$. By means of $m(r)$ from (3.1) and $p_{r}$ from (2.2), one obtains $g_{tt} = -1 + 2ar$. 
   
 For a BH of mass $M$, $a$ - being constant everywhere inside - should equal the surface value. Our choice is to take for $a$ the surface gravity, namely $a = 1/4M$ (it equals the modulus of the radial acceleration obtained from $a^{b} = u^{a}\nabla_{a}u^{b} = (0, -a, 0, 0)$ on the BH horizon. We must remind that $a^{r} = - a < 0$ is the acceleration needed to keep a test particle fixed at $r = const.$ in the interior, otherwise it will go away from the center. The behavior is a consequence of the repulsive nature of the inner dark energy). With $m(r)$ from (3.1), one obtains $g_{tt} = -1 + 2ar = -1 + r/2M$. Consequently, the BH inner metric appears as 
\begin{equation}
 ds^{2} = -(1 - \frac{r}{2M})~ dt^{2} + (1 - \frac{r}{2M})^{-1} dr^{2} + r^{2} d \Omega^{2},~~~r < 2M ,
\label{3.4}
\end{equation}
 with a horizon at $r = 2M$. Being a particular case of the metric (2.1), the above spacetime is an exact solution of Einstein's equation, with the stress tensor (2.2) on its r.h.s. In addition, it is curved and the Kretschmann scalar $K = 2/M^{2}r^{2}$ is divergent at the origin $r = 0$.
 
  We stress again that the exterior metric is obtained using a Newtonian condition at infinity that cannot be applied in the inner part (that is the reason why, in our view, we may not simply replace $r$ with $t$ and vice versa when the Schwarzschild horizon $r = 2M$ is crossed and the inner geometry would become non-static (see \cite{RB, DLC, HC2, DV})). At $r = 2M$, the energy density acquires the value $\rho_{H} = a^{2}/ \pi$, namely $\rho_{H}$ is proportional to the intensity of the field squared \cite{HC3}, as for the electromagnetic field. Moreover, at the horizon the mass $m(r)$ from (3.1) becomes $M$, as expected.
  
 \section{Israel junction conditions}
 The inner metric (3.4) matches the Schwarzschild outer metric 
 \begin{equation}
 ds^{2} = - (1 - \frac{2M}{r}) dt^{2} + (1 - \frac{2M}{r})^{-1} dr^{2} + r^{2} d \Omega^{2},~~~r > 2M
\label{4.1}
\end{equation}
at the interface $r = 2M$ (we note here that $(1 - r/2M)$ may be directly obtained from $(1 - 2M/r)$ by means of an inversion $r \rightarrow (2M)^{2}/r$). We further stress that the norm of the timelike Killing vector does not change its sign when the horizon is crossed, having a single zero at $r = 2M$ (the horizon is nonextremal due to the nonzero surface gravity). The function $-g_{tt}(r), ~r > 0$ is continuous at $r = 2M$ but not differentiable and the norm should not flip from the outside to the inside.

The second Israel junction condition must be satisfied, namely the jump of the extrinsic curvature when the matching surface $r = 2M$ is crossed, should equal the surface stress tensor. To avoid a gluing process on a null surface we take advantage of the Myung prescription \cite{MK} and replace the horizon $r = 2M$ with the so-called ''stretched horizon'', located not at $r = 2M + l_{P}^{2}/2M$ (as he did), where $l_{P}$ is the Planck length (see also \cite{PW}), but at  $r = 2M + l^{2}/2M$, where $l^{2}/2M \equiv \epsilon << 2M$. $l$ is related to the proper distance $\rho$ from the horizon \cite{GP, EB, HC4, SU}, that results from (4.1) by means of the transformation
 \begin{equation}
 d \rho = \frac{dr}{\sqrt{1 - \frac{2M}{r}}} \approx \sqrt{\frac{2M}{r - 2M}} dr.~~~(r \approx 2M,~~r > 2M).
\label{4.2}
\end{equation}
One obtains $\rho = 2\sqrt{2M (r - 2M)} = 2\sqrt{2M\epsilon} = 2l$. 

Being close to the horizon $r = 2M$, we replace $1 - r/2M$ with $1 - (r - \epsilon)/2M$ in (3.4) and $1 - 2M/r$ with $1 - 2M/(r + \epsilon)$ in (4.1). Once the matching conditions are written, we finally take the limit $r \rightarrow 2M$ and then neglect $\epsilon/2M = (l/2M)^{2}$ with respect to unity. 

As far as the second junction condition is concerned, we shall use the Kolekar et al. \cite{KKP} expression of the extrinsic curvature tensor
 \begin{equation}
 K_{\mu \nu} = -\frac{f'}{2\sqrt{f}}u_{\mu}u_{\nu} + \frac{\sqrt{f}}{r}q_{\mu \nu},~~~K = h^{\mu \nu}K_{\mu \nu} = \frac{f'}{2\sqrt{f}} + \frac{2\sqrt{f}}{r}
\label{4.3}
\end{equation}
for the metric
 \begin{equation}
   ds^{2} = - f(r) dt^{2} + f^{-1}(r) dr^{2} + r^{2} d \Omega^{2}.
\label{4.4}
\end{equation}
We have here $\mu, \nu = 0, 2, 3, f' = df/dr,~ u_{\mu} = (\sqrt{f}, 0, 0) $ is the normal to $t = const.$ hypersurface, $h_{\mu \nu} = (g_{ab} - n_{a}n_{b})\delta_{\mu}^{a} \delta_{\nu}^{b}$ represents the induced metric on $r = const.$ surface, $n_{a} = (0, 1/\sqrt{f}, 0, 0)$ is its normal vector and $q_{\mu \nu} = h_{\mu \nu} + u_{\mu}u_{\nu}$ - the induced metric on the level surfaces of constant time hypersurfaces \cite{KKP}.

The Lanczos equation gives the $r = const.$ surface stress tensor $S_{\mu \nu}$ 
  \begin{equation}
 [- K_{\mu \nu} + h_{\mu \nu}K] = 8\pi S_{\mu \nu}
\label{4.5}
\end{equation}
with its perfect fluid form
  \begin{equation}
  S_{\mu \nu} = \rho_{S}u_{\mu}u_{\nu} + p_{S} q_{\mu \nu}.
\label{4.6}
\end{equation}
$\rho_{S}$ is the surface energy on $r = 2M$, $p_{S}$ is the surface pressure and $[K_{\mu \nu}] = K_{\mu \nu}^{+} - K_{\mu \nu}^{-}$ is the jump of the extrinsic curvature when the thin layer of thickness $(2M + \epsilon) - (2M - \epsilon) = 2\epsilon$ is crossed. With $f_{+} = 1 - 2M/(r + \epsilon)$ and $f_{-} = 1 - (r - \epsilon)/2M$, Eqs. (4.3) yield
  \begin{equation}
  \begin{split}
 [- K_{tt} + h_{tt}K] = - \frac{2}{r} \left(\left(1 - \frac{2M}{r + \epsilon}\right)^{3/2} - \left(1 - \frac{r - \epsilon}{2M}\right)^{3/2}\right)\\
 [- K_{\theta \theta} + h_{\theta \theta}K] = + r \left(\left(1 - \frac{2M}{r + \epsilon}\right)^{1/2} - \left(1 - \frac{r - \epsilon}{2M}\right)^{1/2}\right)\\
 + Mr^{2}\left(\frac{1}{(r +  \epsilon)^{2}} \left(1 - \frac{2M}{r + \epsilon}\right)^{-1/2} + \frac{1}{4M^{2}} \left(1 - \frac{r - \epsilon}{2M}\right)^{-1/2}\right),
\label{4.7}
\end{split}
\end{equation}
evaluated at $r = 2M$. If one neglects $\epsilon/2M$ w.r.t. unity, one finds from (4.6) and (4.7) that
  \begin{equation}
 S_{tt} = 0,~~~8\pi S_{\theta \theta} = 2M \sqrt{\frac{2M}{\epsilon}},~~~S_{\phi \phi} = sin^{2}\theta  ~S_{\theta \theta}
\label{4.8}
\end{equation}
Keeping now in mind that we have chosen $\epsilon = l^{2}/2M$, Eqs. (4.8) lead to 
  \begin{equation}
  \rho_{S} = 0,~~~p_{S} = \frac{1}{8\pi l}
\label{4.9}
\end{equation}
in the approximation used. A null surface energy $\rho_{S}$ has also been obtained by Kolekar et al. \cite{KKP} in their study of the entropy of spherically symmetric gravitational shells on the verge of forming a BH. $p_{S}$ diverges, of course, when $\epsilon$ tends to zero. A good choice to avoid that would be to take $l = l_{P}$, as Myung did. Quantum geometry does not give a better choice. We would also like to stress that the surface pressure $p_{S}$ contributes to the gravitational energy as the integrand in the expression of the Tolman-Komar energy \cite{TP3} contains both the energy density and pressures.

To find the explicit form of the induced metric on the horizon $r = 2M$, we take firstly in (3.4) or (4.1) a surface $\Sigma$ of constant $r$ and then replace $r$ with $2M \pm \epsilon$, to have
  \begin{equation}
  ds^{2} = - \frac{\epsilon}{2M} dt^{2} + 4M^{2} (1 \pm \frac{\epsilon}{2M})^{2} d \Omega^{2}. 
\label{4.10}
\end{equation}
Taking the limit $\epsilon \rightarrow 0$, Eq. (4.10) yields on the horizon 
  \begin{equation}
  ds^{2}|_{H} =  4M^{2} (d\theta^{2} + sin^{2}\theta d\phi^{2}). 
\label{4.11}
\end{equation}
A similar result has been obtained by Gourgoulhon and Jaramillo \cite{GJ} (their Eq. (4.86)), in the study on null hypersurfaces. (4.11) is a two-dimensional surface and that is in accordance with the well-known idea that there is no time on the BH horizon (everything, even light, stands still on it). An extra support for this interpretation comes from the fact that $S_{tt} =0,~\rho_{S} = 0$ on the horizon.

For the normal accelerations to the stretched horizons we have, respectively
  \begin{equation}
  \begin{split}
  a^{+} \equiv n_{b}^{+} a^{b}_{+} = \frac{M}{(r +  \epsilon)^{2}} \left(1 - \frac{2M}{r + \epsilon}\right)^{-1/2}\\
  a^{-} \equiv n_{b}^{-} a^{b}_{-} = - \frac{1}{4M} \left(1 - \frac{r - \epsilon}{2M}\right)^{-1/2}.
\label{4.12}
\end{split}
\end{equation}
$n_{b}^{\pm}$ are expressed in terms of $f_{+}$ and, respectively, $f_{-}$, and the accelerations $a^{b} = u^{a}\nabla_{a}u^{b}$ are obtained from $u^{b} = (\sqrt{f}, 0, 0, 0)$.  
With $r = 2M$, one obtains
  \begin{equation}
a^{+} = \frac{1}{4M} \sqrt{\frac{2M}{\epsilon}},~~~a^{-} = - \frac{1}{4M} \sqrt{\frac{2M}{\epsilon}}  
\label{4.13}
\end{equation}
When $\epsilon = l^{2}/2M$ is substituted in (4.13), we get $a^{+} = - a^{-} = 1/2l = 1/\rho$. What we carried out here was to remove the divergence of the proper acceleration at the Schwarzschild horizon by means of the stretched horizons, located close to the event horizon, at $r = 2M \pm l^{2}/2M$. We so rendered the acceleration finite, which reaches half of its Planck value when $l = l_{P}$. The discontinuity of $K_{\mu \nu}$ implies a discontinuity of the normal accelerations \cite{BGG, IS}. The fact that $a^{-} < 0 $ is rooted from the repulsive nature of the dark energy from the BH interior (characterized by negative pressures).

To test whether the assumed inner BH metric (3.4) is valid, we may consider a Schwarzschild observer $O$, freely falling radially into the BH. He/she finds nothing special when the horizon $r = 2M$ is crossed. Once inside, $O$ will move according to a timelike inward geodesic obtained from (3.4)
  \begin{equation}
  r(t) = 2M \left(1 - \frac{E^{2}}{cosh^{2} \frac{t}{4M}}\right) ,
\label{4.14}
\end{equation}
under suitable initial conditions ($0 < E < 1$ is a constant related to observer's energy and $t\in(-\infty, \infty)$). Notice that there is a $r_{min} = 2M(1 - E^{2})$, as if $O$ were rejected by a central core, due to the repulsive character of the inner dark energy. At $r = r_{min}, O$ sends a light signal to the center $r = 0$ , that will travel according to the ingoing null radial geodesic
  \begin{equation}
  r_{n}(t) = 2M \left(1 - e^{- \frac{|t|}{2M}} \right) ,~~~t \in(- \infty, \infty),
\label{4.15}
\end{equation}
where the index ''n'' refers to ''null''. Once the signal reaches the central singularity, it will become outgoing and will meet the observer $O$ after some moment $t_{0}$ obtained from $r(t_{0}) = r_{n}(t_{0})$, namely $t_{0} = - ln(2E - 1) > 0, (1/2 < E < 1)$. Measuring $t_{0}$ gives $O$ the possibility to check the validity of the geometry (3.4). It is worth noting that a static observer from the Schwarzschild region cannot get informations from $O$ because the outgoing light signal will never cross the horizon. We also mention that all geodesic are complete since they do not terminate at the singularity but continue to infinite time.

\section{Interior entropy} 
 As we already have observed, Davidson and Gurwich \cite{DG2} have obtained a similar expression for $m(r)$, proportional to the radius squared, in their model on holographic entropy inside a black hole. They showed that the inner metric has no signature flip at the transition surface $r = 2M$. The same property is valid for the couple of geometries (3.4) - (4.1).
  
 Our purpose is now to compute the entropy $S(r)$ inside the BH. We shall use their prescription, taking advantage of the fact that our metric (3.4) is of Rindler type. Therefore, we recover the Hawking temperature $T_{H} = 1/8 \pi M$ from this Rindler structure \cite{DG2}. Keeping in mind that we must take into account the inner pressures, the energy balance should be written as
 \begin{equation}
 \delta m(r) = T_{H} \delta S(r) - (p_{r} - p_{\bot}) \delta V,
\label{5.1}
\end{equation}
when $m(r)$ undergoes a tiny variation $\delta m(r)$. We have $m(r) = r^{2}/4M$, $p_{r} = 2p_{\bot} = -1/8 \pi Mr$ and $V = 4 \pi r^{3}/3$. One obtains 
\begin{equation}
S(r) = 8 \pi M \frac{r^{2}}{8M} =  \pi r^{2},
\label{5.2}
\end{equation}
independent of the BH mass $M$. 

Let us remark that the entropy and volume increase in (5.1) comes from a tiny augment of $r$, after adding some mass $\delta m(r)$. The expression (5.2) for the entropy corresponds exactly to that one obtained by the authors of \cite{DG2}. When $r = 2M$, the Bekenstein-Hawking formula $S = 4 \pi M^{2}$ is recovered. 

From (3.1) and (5.2) we have $S(r)/m(r) = 4 \pi M = S/M$ where $S$ is the BH entropy. We also note that $M = 2TS$, as Padmanabhan \cite{TP2} obtained before, keeping track of the pressure term in the expression of the first law. As Davidson and Garwich already noticed, the entropy $S(r)$ of any BH inner sphere is maximally packed, being unaffected by the outer layers. In other words, the degrees of freedom are spread in the whole BH interior (see also \cite{EB2}).

\section{Penrose diagram}
Let us now study the global structure of the spacetime (3.4). We have a singularity at $r = 0$ (the Kretschmann scalar diverges there) and a horizon at $r = 2M$. Keeping in mind that the Weyl tensor vanishes in the geometry (3.4), there will be a coordinate system where it is conformally flat. This feature will prove to be useful in drawing the conformal diagram.

We follow Kiselev \cite{VVK} method for to transform the metric (3.4) in a conformally flat form. With the help of the coordinate transformation  $\chi = (-1/2) ln(1 - r/2M)$, it may be expressed as 
\begin{equation}
ds^{2} = \frac{1}{a^{2}} e^{-2\chi} (- a^{2} dt^{2} + d\chi^{2} + sinh^{2}\chi d \Omega^{2})
\label{6.1}
\end{equation}
where $a = 1/4M$ is the horizon surface gravity. Kiselev further makes the Fock transformation from the hyperbolic coordinates to the flat ones
\begin{equation}
\tau = e^{a t} cosh\chi,~~~\rho = e^{a t} sinh\chi,
\label{6.2}
\end{equation}
to find
\begin{equation}
ds^{2} = \frac{1}{a^{2}(\tau + \rho)^{2}} (- d\tau^{2} + d\rho^{2} + \rho^{2} d \Omega^{2}),
\label{6.3}
\end{equation}
which has the desired conformally flat structure. We have above $\chi > 0,~\tau > 0$ and $0 < \rho <\tau$. The last equation reveals $\rho = 0$ as a singular point (it corresponds to $r = 0$ in the original metric). In addition, the condition $\rho + \tau \rightarrow \infty$ is equivalent to $r \rightarrow 2M$ in (3.4) (we note that (6.2) is not spatially homogeneous due to the $\rho$-dependence of the conformal factor). Hence, in contrast to the conformally flat de Sitter universe the geometry (6.3) does not allow a FRW type evolution \cite{VVK}. 

We are now in a position to represent the conformal diagram of the spacetime (3.4), taking advantage of its conformally flat form from (6.3). Keeping in mind that the causal structure of a spacetime is unchanged by a conformal compactification, we drop the conformal factor $1/a^{2}(\tau + \rho)^{2}$ and obtain the Minkowski metric in spherical coordinates. We further follow up the standard procedure \cite{PD, PT, HC6}, suppress the angular coordinates and define
 \begin{equation}
 T = arctan(\tau + \rho) + arctan(\tau - \rho),~~~R = arctan(\tau + \rho) - arctan(\tau - \rho),
\label{6.4}
\end{equation}
 with $0 < T < \pi,~0 < R < \pi/2$. The corresponding Penrose diagram is depicted in Fig.1. We observe that the $r = 0$ singularity is timelike because no signature flip occurs when the horizon $r = 2M$ is crossed. The boundaries of the diagram are the ''past null horizon'' (PH) and, respectively, ''the future null horizon'' (FH), both located at $r = 2M$. A light ray emerging from PH bounces at the singularity $r = 0$ and then reaches FH after infinite time $t$, as can be seen from the null geodesics (4.15). 
 
 A free massive particle emerging from $i^{-}$ is rejected by the singularity at a distance depending on the initial conditions (as per Eq. (4.14) for the timelike geodesics) and then approach $i^{+}$ at $T = \pi$ (the line $T = \pi/2$ corresponds to the time $t = 0$).

\section{Conclusions} 
To summarize, we proposed in this Letter a model for the BH interior leading to several similar conclusions with that of Davidson and Gurwich. In spite of the central singularity, the mass $m(r)$ is finite everywhere inside and the entropy function $S(r)$ is maximal for any radius $r$, that is it equals the BH entropy. The metric is continuous at the horizon and the singularity at the origin contributes nothing to the expressions of $m(r)$ and $S(r)$. We have also studied the Israel junction conditions on the stretched horizon so avoiding the matching process on the null surface $r = 2M$. A conformal diagram is represented for $r < 2M$ in fig. 1, with a timelike singularity at $r = 0$ where the null particles are repelled because of the repulsive character of the inner gravitational field.

\end{document}